\documentclass[sigconf,nonacm]{acmart}

\usepackage{balance} 

\usepackage{booktabs}
\usepackage{tabularx}
\usepackage[most]{tcolorbox}
\definecolor{promptframe}{HTML}{E47D70}
\definecolor{sysbg}{HTML}{EAF1F9}
\definecolor{sysaccent}{HTML}{6A9AD0}
\definecolor{userbg}{HTML}{F0F6E8}
\definecolor{useraccent}{HTML}{7FAE4F}
\newtcolorbox{promptbox}[1]{enhanced, breakable=false,
  colback=white, colframe=promptframe, boxrule=0.6pt, arc=1.5mm,
  left=1.2mm, right=1.2mm, top=1.2mm, bottom=1.2mm,
  title={#1}, fonttitle=\sffamily\bfseries\small, coltitle=black,
  colbacktitle=promptframe, toptitle=0.5mm, bottomtitle=0.5mm}
\newtcolorbox{rolebox}[3]{enhanced, sharp corners,
  colback=#2, colframe=#3, boxrule=0pt, leftrule=2.2pt,
  left=1.6mm, right=1.4mm, top=1mm, bottom=1mm,
  fontupper=\small, before upper={\raggedright{\sffamily\bfseries\footnotesize\color{#3}#1}\par\smallskip}}
\newcommand{\promptfield}[1]{\par\smallskip\noindent{\color{useraccent}$\triangleright$}~\texttt{"#1"}:~}
\usepackage{fvextra}
\usepackage{placeins} 
\DefineVerbatimEnvironment{promptverb}{Verbatim}{breaklines, fontsize=\footnotesize}

\setcopyright{none}
\title[Agentic-Priors]{Agentic-Priors:
  Using Large Language Models to Address the Cold Start Problem in
Preference Elicitation}

\author{Bruno Arcanjo}
\affiliation{
  \institution{University of Southampton}
  \city{Southampton}
\country{United Kingdom}}
\email{brqa1a24@soton.ac.uk}

\author{Zhaoxing Li}
\affiliation{
  \institution{University of Southampton}
  \city{Southampton}
\country{United Kingdom}}
\email{zhaoxing.Li@soton.ac.uk}

\author{Sebastian Stein}
\affiliation{
  \institution{University of Southampton}
  \city{Southampton}
\country{United Kingdom}}
\email{ss2@soton.ac.uk}

\begin{abstract}
An agent that learns a user's preferences from their choices faces a
cold start problem: before any choices are observed, it must rely solely on a
prior. Informative priors are usually fitted to historical choice data, which new tasks and deployments lack. To address this, we
present Agentic-Priors, which lets agentic systems use the implicit knowledge of large language models (LLMs) instead. Specifically, an LLM states a new user's trade-offs in interpretable terms, such as willingness to pay or value of time, which we map to a prior over multinomial logit coefficients. This  prior is then refined by a Bayesian learner as choices are observed. 
Across seven choice datasets,
Agentic-Priors improves cold start accuracy over an uninformed prior by 5--40 percentage points. It also outperforms AutoElicit, a state-of-the-art LLM prior-elicitation method, on three datasets and loses on one. It is up to 18 points
more accurate than in-context learning, and
never significantly worse. Against hierarchical Bayes models fitted to surveys of increasing size, our approach is worth roughly 6--40 surveyed users, in terms of log-loss, so its value is concentrated in a deployment's first users.
\end{abstract}


\newcommand{\BibTeX}{\rm B\kern-.05em{\sc i\kern-.025em b}\kern-.08em\TeX}

\begin{document}


\maketitle


\section{Introduction}

For an agent to make personalised recommendations, or even autonomous decisions, it must estimate a
user's preferences. A standard approach observes the user's choices
and updates a utility model by Bayesian
inference~\cite{chajewska2000adaptive,boutilier2002pomdp}. Before the first
choice, however, the agent can rely only on its prior, and a poor prior
means poor early recommendations and slower learning. This is known as the
cold start problem.

The usual remedy is population data. A hierarchical Bayes model (HBM)
fitted to other users' choices on the same task gives an informative
prior~\cite{allenby1998marketing}, and conditioning it on user
characteristics makes that prior user-specific~\cite{campigotto2017favour}. But this needs a choice dataset for the task at hand,
which a new product, service or deployment does not have. Without one,
the agent falls back on an uninformed prior, which says nothing
about what the user prefers.

Large language models (LLMs) encode broad knowledge about how people
make choices~\cite{argyle2023out,goli2024frontiers}, which suggests a third option: using this implicit knowledge to derive more informative priors. Specifically, in our approach, an LLM reasons about how a user would trade off the attributes of a choice, and
states those trade-offs in a form a choice model can use. This can then be used to provide relevant recommendations or make autonomous actions on the user's behalf. Here, the difficulty is that utility coefficients
have no natural units~\cite{train2009discrete}, so an LLM has nothing to anchor their magnitude
to. We instead ask for trade-offs on a reference scale it can reason
about, such as how much the user would pay to save an hour of travel,
and convert these into coefficients.

\begin{figure}[t]
  \centering
  \includegraphics[width=0.8\columnwidth]{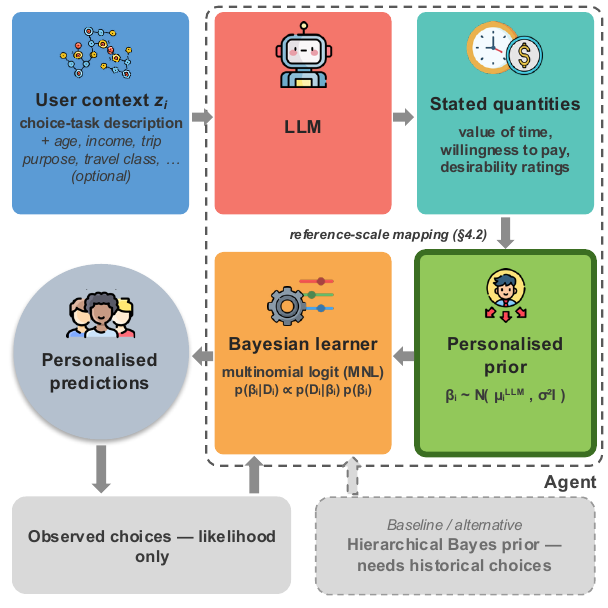}
  \Description{A flow diagram of the Agentic-Priors pipeline.}
  \caption{Agentic-Priors overview. An LLM, given the choice task
  and (optionally) the user's characteristics $z_i$, states
  interpretable quantities such as willingness to pay. These are mapped
  to multinomial logit (MNL) coefficients that form a personalised prior
  $\beta_i \sim \mathcal{N}(\mu_i^{\mathrm{LLM}}, \sigma^2 I_K)$, which a
  Bayesian learner updates as choices are observed. Unlike a
  hierarchical Bayes prior, it needs no historical choice data.}
  \label{fig::pipeline}
\end{figure}

Against this background, we introduce \textbf{Agentic-Priors}, which elicits a user-specific
prior from a single LLM call. The LLM is given a description of the choice task and, optionally, whatever is known about the user, such as their demographics or other relevant context (e.g.~the purpose of a trip). Agentic-Priors requires no historical choice data, and
then hands the prior to a Bayesian learner that updates it as real
choices arrive (Figure~\ref{fig::pipeline}), e.g.~when a user chooses from a set of presented options. We evaluate it on seven stated- and revealed-choice
datasets from transport, vehicle purchase, consumer goods and finance,
against an uninformed prior, an LLM that predicts choices directly using in-context learning~\cite{brown2020language}, and AutoElicit~\cite{capstick2025autoelicit}, a state-of-the-art approach that uses LLMs as priors for general machine learning problems. We also compare our approach to HBM~\cite{allenby1998marketing,rossi2003bayesian}, which is widely used in choice modelling when detailed historical population data is available.

Agentic-Priors improves cold start accuracy over an uninformed prior on all
seven datasets, by 5 to 40 percentage points. In accuracy, it
outperforms AutoElicit on three datasets, loses on one and is
inconclusive on the remaining three. Predicting choices directly with the LLM, with past
choices given in context, is never significantly better than eliciting
a prior and updating it, and is up to 18 percentage points less accurate. To
measure what the prior is worth in data, we compare it with HBMs
fitted to surveys of increasing size: in log-loss it is worth between
about six and about forty surveyed users, depending on the
dataset, and in accuracy at most about six on all datasets but one.
Its value is therefore real but confined to a deployment's first
users. Moreover, on most datasets a prior elicited from the task description alone, without any information about the user, is worth almost as much: the value lies mainly in the LLM's knowledge of the task.

Our contributions are:
\begin{itemize}
  \item \textbf{Agentic-Priors}, a method that turns an LLM's
    reasoning about a user into a prior over choice-model coefficients. This uses a reference scale that gives the elicited quantities meaningful units and produces a choice model prior that can be updated using Bayesian learning (Section~\ref{sec::llm-priors}).
  \item An \textbf{evaluation} on seven real-world datasets against a range of state-of-the-art baselines that use no data, population data, or in-context learning (Section~\ref{sec::results}).
  \item An \textbf{exchange rate} that expresses the value of an LLM
    prior as the number of surveyed users it is worth, a concrete basis for deciding when one is worth using (Section~\ref{sec::res_exchange}).
\end{itemize}

\section{Related Work}

An agent that recommends or decides on a user's behalf must
estimate their preferences from few interactions---the cold start
problem of recommender systems~\cite{schein2002methods}. Research on
preference elicitation has largely optimised \emph{which} comparisons
to present: decision-theoretic formulations select queries that
reduce utility uncertainty or minimax regret
\cite{chajewska2000adaptive,boutilier2002pomdp,braziunas2007minimax,zintgraf2018ordered},
and adaptive conjoint and efficient choice designs pursue the same
goal through experimental design
\cite{toubia2004polyhedral,saure2019ellipsoidal,rose2009constructing}.
Every such method, however, proceeds from what the agent already
believes before the first question is asked---and that belief alone
determines its behaviour while no choices have yet been observed. In contrast, we here study the complementary lever: the user-specific information available before any interaction.

Random-utility models make that belief explicit and interpretable,
representing a user's preferences as a coefficient vector over
attribute trade-offs \cite{mcfadden2000mixed,train2009discrete}.
Conventionally, the distribution over those coefficients is estimated
from a population of earlier users, with mixed logit and
hierarchical Bayes capturing how preferences vary across individuals
\cite{allenby1998marketing}. Closest to our setting, FAVOUR maps
sociodemographic profiles to population-informed priors for
personalised routing and then refines them by Bayesian learning
\cite{campigotto2017favour}. This pipeline is similar to ours, but differs in where the prior comes from. Recent personalisation methods
likewise learn shared preference structure from prior users before
adapting to a new one
\cite{poddar2024vpl,shenfeld2025pref,bose2026pep}. All provide
informative starting points, but all require a historical choice
dataset from comparable users performing a comparable
task---assumptions we make explicit in
Section~\ref{sec::conventional_priors}, and whose cost motivates this work.

LLMs offer a different source for the same object. One line of work elicits
priors for Bayesian models directly from an LLM and examines how well
its stated uncertainty is calibrated
\cite{capstick2025autoelicit,renda2026openestimate}, though the
elicited prior is one per task rather than one per user, and the work considers general machine learning tasks rather than choice models; hence, we adapt this approach to choice data, with and without user conditioning, as baselines. A
second line of work simulates users or augments human choice data, including
individual-level mimicry conditioned on socio-economic microdata,
with mixed evidence on behavioural fidelity
\cite{argyle2023out,bisbee2024synthetic,goli2024frontiers,wang2026market,kinzinger2026synthetic};
these yield simulated answers rather than a distribution over
preference coefficients. LLMs given stated preferences are also
competitive recommenders near cold start~\cite{sanner2023large}, but
return recommendations, not a model of trade-offs. A third line places LLM knowledge inside
sequential decision loops: CBLI warm-starts bandits with synthetic
preferences and subsequent work identifies negative transfer when
those preferences are misaligned
\cite{alamdari2024jump,bayley2026jumpstart}, while OPEN combines LLMs
with Bayesian experimental design to select informative queries
\cite{handa2024open}. There the knowledge enters as synthetic
training data for a black-box reward model, not as a prior over
interpretable coefficients.

We combine these perspectives: in Agentic-Priors, an LLM supplies a
prior over the coefficients of an interpretable choice model,
conditioned on whatever is known about the user and requiring no
historical choice data. The user's own choices enter afterwards, through
the likelihood alone (Section~\ref{sec::bayes_update}). This isolates
what encoded world knowledge contributes at cold start, and
allows us to ask---against priors that do draw on historical
choices---whether that contribution persists as evidence accumulates.

\section{Preference Modelling}

This section lays the necessary groundwork for our method Agentic-Priors presented in Section~\ref{sec::llm-priors} by
formulating the preference learning and elicitation problem,
detailing notation and assumptions. We start from the discrete choice
problem and how user preferences naturally arise from the
formulation. We follow with how Bayesian learning is used to update
preferences as new choices are observed. Finally, we describe the
conventional priors, with and without population data.

\subsection{Discrete Choice Problem}

In a discrete-choice setting~\cite{benakiva1985discrete,train2009discrete}, each user $i$ completes a
sequence of choice tasks $t = 1, 2, \dots$, in each choosing one of $J$
alternatives described by $K$ attributes whose values differ between
tasks (e.g.~A, B or C first, then D, E or F). Alternative $j$ in task
$t$ has attribute vector $x_{ijt} \in \mathbb{R}^K$, and
$X_{it} = (x_{i1t}, \dots, x_{iJt})$.

The utility $U_{ijt}$ that user $i$ derives from alternative $j$ in task $t$ is
represented by the linear combination

\begin{equation}
  U_{ijt} = x_{ijt}^T\beta_i + \epsilon_{ijt}
\end{equation}
where $\beta_i$ is the coefficient vector representing user
$i$'s preferences, the same in every task, and $\epsilon_{ijt}$ is an error term, assumed to be i.i.d.~Gumbel, accounting for
unobserved randomness. The MNL model~\cite[Ch.~3]{train2009discrete} then calculates the
probability that user $i$ chooses alternative $j$ in task $t$ as follows

\begin{equation}
  P(y_{it} = j \mid X_{it}, \beta_i) =
  \frac{\exp({x_{ijt}^T\beta_i)}}{\sum_{l=1}^{J}{\exp(x_{ilt}^T\beta_i)}}
\end{equation}
where $y_{it} \in \{1, \dots, J\}$ is user $i$'s choice in task $t$.

In this work, the objective of preference learning is to find an
estimate $\hat{\beta}_i$ that best approximates the true, latent
preference vector $\beta_i$. We make an explicit distinction between
two information sources for estimating $\beta_i$: historical population-level
data and the user's observed choices.

\subsection{Bayesian Updates}
\label{sec::bayes_update}

Before observing any choices, we hold a prior belief $p(\beta_i)$ of
user $i$'s preferences. If relevant population-level data is
available, $p(\beta_i)$ can be estimated from it; otherwise, an
uninformed prior is used (Section~\ref{sec::conventional_priors}).

After observing user $i$'s first $T$ choices, $D_i = \{(X_{it}, y_{it})\}_{t=1}^{T}$ (so $T = |D_i|$), $p(\beta_i)$ is updated by Bayes' rule,
\begin{equation}
  p(\beta_i \mid D_i) \propto p(D_i \mid \beta_i)\,p(\beta_i),
\end{equation}
where $p(\beta_i \mid D_i)$ is the posterior and $p(D_i \mid \beta_i)$
is the MNL likelihood which, assuming the $T$ observed choices are
conditionally independent given $\beta_i$, is
$p(D_i \mid \beta_i) = \prod_{t=1}^{T}P(y_{it} \mid X_{it}, \beta_i)$.

We define a cold start at $T=0$ where $D_i = \emptyset$, i.e.~there
are yet no observed choices for user $i$. Consequently,

\begin{equation}
  p(\beta_i \mid D_i=\emptyset) = p(\beta_i)
  \label{eq::cold_start_prior}
\end{equation}

Therefore, in a cold start scenario, the user-specific
preference information comes entirely from the prior. Thus, methods
for estimating the starting $p(\beta_i)$ become particularly
important for addressing the cold start problem.

\subsection{Conventional Priors}
\label{sec::conventional_priors}

A historical choice dataset gives an informative prior when its
users share the new user's population of preferences and
performed a comparable task, with compatible attributes and utility
model. The standard approach is a hierarchical model,
$\beta_i \sim \mathcal{N}(\mu^{\mathrm{pop}}, \Sigma^{\mathrm{pop}})$,
fitted jointly to all historical users' choices, where
$\mu^{\mathrm{pop}}$ may depend on user characteristics\cite{allenby1998marketing}. A new user then starts from
$\beta_i \sim \mathcal{N}(\mu^{\mathrm{pop}}, \Sigma^{\mathrm{pop}})$:
$\mu^{\mathrm{pop}}$ sets where preferences are centred, and
$\Sigma^{\mathrm{pop}}$ how much they vary across users and how
the coefficients covary. Without such a dataset, one resorts to an
uninformed prior, $\beta_i \sim \mathcal{N}(0, \sigma^2 I)$, which
carries no user-specific information. In the next section, we
replace the source of the prior with the knowledge encoded in an LLM,
which needs no historical data.

\section{Agentic-Priors for Preference Elicitation}
\label{sec::llm-priors}

Agentic-Priors elicits a prior over MNL
coefficients from an LLM, given only a description of the
choice task and whatever is known about the user, $z_i$, e.g.~demographics or additional contextual information (possibly nothing). It requires no historical choice data. This section describes what the LLM is
given (Section~\ref{sec::prompt}), how its answers are mapped to MNL
coefficients (Section~\ref{sec::mon_eq_fmrk}), and how a prior is formed
around them (Section~\ref{sec::llm_prior_construction}).

\subsection{LLM Input}
\label{sec::prompt}

The LLM receives two messages via a single prompt. The \emph{system message} is fixed per
task: it sets the LLM's role, names the alternatives, states which
attributes reduce utility, and requires a JSON reply. The \emph{user
message} is specific to the user: it describes $z_i$ in a short
natural-language profile, lists any choices the user has already
made (in our cold start setting this is always empty), and asks for each quantity on
the task's reference scale (Section~\ref{sec::mon_eq_fmrk}), with a
one-line definition and sign convention for each. We give the LLM
only the user information each dataset provides
(Table~\ref{tab::dataset_characteristics}). When $z_i$ is omitted, the
profile is replaced by a generic description (e.g.\ ``a Swiss intercity
traveller''), so the LLM relies on its knowledge of the task alone.
Figure~\ref{fig::prompt}
shows the Swissmetro messages for one user (datasets: Section~\ref{sec::datasets}).

\begin{figure}[t]
\begin{promptbox}{LLM input \textperiodcentered\ Swissmetro user}
\begin{rolebox}{System}{sysbg}{sysaccent}
You are a transport-choice analyst estimating one
traveller's preferences for a discrete-choice model of Train vs
Swissmetro (a fast mag-lev) vs Car. Travel time and cost reduce
utility. Reply with ONLY a JSON object, no prose.
\end{rolebox}
\vspace{-1mm}
\begin{rolebox}{User}{userbg}{useraccent}
Traveller profile: a woman aged 39--54, annual income
CHF 50,000--100,000, travelling for commuting, who normally travels
second class, with no luggage, paying their own fare.
No past choices are available for this traveller.
Estimate this traveller's preferences as JSON with these numeric
fields (money is the unit throughout):
\promptfield{value\_of\_time\_chf\_per\_hour} CHF they would pay to save one hour of
travel time (positive);
\promptfield{train\_vs\_swissmetro\_chf} the extra CHF they would pay for Train
over Swissmetro if the two had identical travel time and cost, i.e.\
their intrinsic mode preference only (negative if they would still
prefer Swissmetro);
\promptfield{car\_vs\_swissmetro\_chf} the same for Car vs Swissmetro.
\end{rolebox}
\end{promptbox}
\caption{LLM input for one Swissmetro user (abridged: definitions
shortened). Prompts for all datasets are in
Appendix~\ref{app::prompts}.}
\label{fig::prompt}
\Description{A boxed example of the two messages sent to the LLM for one
Swissmetro user: a system message setting the analyst role, and a
user message with their profile and the three monetary
quantities to estimate.}
\end{figure}

\subsection{Reference-Scale Framework}
\label{sec::mon_eq_fmrk}

Guided by the prompt, the LLM expresses each preference as a
quantity $a_{ik}$ on a common reference scale, so that preferences over
different attributes are directly comparable. A reference coefficient
$c_{\mathrm{ref}}$ fixes the utility value of one unit of that scale, and a
task-specific map converts each answer into a coefficient:

\begin{equation}
\beta^{\mathrm{raw}}_{ik} = \delta_k\, c_{\mathrm{ref}}\,\frac{a_{ik}}{u_k},
\end{equation}

where $u_k$ converts units and $\delta_k \in \{-1, +1\}$ is $-1$ for
attributes the LLM values as reductions (e.g.\ paying to save time)
and $+1$ otherwise.

When the task has a monetary attribute, money is the natural reference
scale: $a_{ik}$ is the user's willingness to pay for (or to avoid)
attribute $k$, and the money attribute is the numeraire, with
$\beta^{\mathrm{raw}}_{\mathrm{num}} = -c_{\mathrm{ref}}$. For example, in Swissmetro,

\begin{equation}
\beta^{\mathrm{raw}}_{\mathrm{time}} = -c_{\mathrm{ref}}\,\frac{\text{VOT}}{60},
\end{equation}

with $\delta_{\mathrm{time}} = -1$, $a_{i,\mathrm{time}} = \text{VOT}$ (the stated value of
time in CHF per hour) and $u_{\mathrm{time}} = 60$, converting CHF per hour into
CHF per minute. Both $\beta^{\mathrm{raw}}_{\mathrm{cost}}$ and $\beta^{\mathrm{raw}}_{\mathrm{time}}$ are negative
because higher cost and longer travel time reduce utility. In the Cars
dataset,

\begin{equation}
\beta^{\mathrm{raw}}_{\mathrm{HEV}} = c_{\mathrm{ref}}\,\frac{\text{WTP}_{\mathrm{HEV}}}{1000},
\end{equation}

with $\delta_{\mathrm{HEV}} = +1$, $a_{i,\mathrm{HEV}} = \text{WTP}_{\mathrm{HEV}}$ (what the
user would pay for a hybrid over an otherwise identical gasoline
vehicle) and $u_{\mathrm{HEV}} = 1000$, since price is measured in thousands of
dollars. $\text{WTP}_{\mathrm{HEV}}$ can be negative, in which case the
coefficient is negative too.

When no attribute has a meaningful monetary value, the LLM instead
rates each attribute level on a bounded desirability scale,
$a_{ik} \in [-100, 100]$, relative to the attribute's baseline level
($\delta_k = +1$, $u_k = 100$), and $c_{\mathrm{ref}}$ is the utility of a
maximally (un)desirable level.

The resulting coefficients must then be expressed on the same scale
as the features used by the discrete-choice model. The choice
attributes are normalised by their design standard deviations,

\begin{equation}
x_{ijtk}^{\mathrm{norm}} = \frac{x_{ijtk}}{s_k},
\end{equation}

where $s_k$ is the standard deviation of attribute $k$ across all
alternatives in all choice tasks of the design. To preserve utility,
the coefficients are scaled accordingly:

\begin{equation}
\beta_i^{\mathrm{scaled}} = D_s\beta_i^{\mathrm{raw}}, \qquad
D_s = \operatorname{diag}(s_1, \dots, s_K).
\end{equation}

Finally, the coefficient vector is divided by the design standard
deviation of the reference attribute, $s_{\mathrm{ref}}$:

\begin{equation}
\beta^{\mathrm{norm}}_i = \frac{\beta_i^{\mathrm{scaled}}}{s_{\mathrm{ref}}},
\end{equation}

where $s_{\mathrm{ref}} = s_{\mathrm{num}}$ when a numeraire exists, which anchors
$|\beta^{\mathrm{norm}}_{\mathrm{num}}| = c_{\mathrm{ref}}$, so that $c_{\mathrm{ref}}$ becomes the
utility of a one-standard-deviation change in the numeraire, and
$s_{\mathrm{ref}} = 1$ otherwise.

\subsection{Prior Construction from a Single Elicitation}
\label{sec::llm_prior_construction}

For each user $i$, we query the LLM once, with $z_i$ when it
is available, under a fixed prompt and decoding configuration, and use
the normalised coefficient vector from Section~\ref{sec::mon_eq_fmrk}
as the prior mean, $\mu_i^{\mathrm{LLM}} = \beta_i^{\mathrm{norm}}$. Unlike
the population priors in Section~\ref{sec::conventional_priors}, this
requires no historical choice data; with $z_i$ it is specific to each user,
and without it every user gets the same prompt.

A single elicitation gives a point estimate but no measure of
uncertainty, so we do not directly derive the covariance from the LLM's
outputs. We use an isotropic covariance with one scale $\sigma$, shared
by all coefficients and fixed across datasets:

\begin{equation}
\beta_i \sim \mathcal{N}\!\left(\mu_i^{\mathrm{LLM}},\, \sigma^2 I_K\right).
\end{equation}

Normalisation by design standard deviations, together with the
reference scale, puts coefficients on a comparable scale across
attributes and datasets, so one shared $\sigma$ is reasonable; it also
keeps the prior broad enough for a few observed choices to move the
posterior (Section~\ref{sec::bayes_update}). We also experimented with asking
the LLM for a plausible range for each quantity and with deriving the
covariance from the spread of repeated elicitations, but both gave
priors far too narrow to learn from: the stated ranges had to be widened
about 16-fold, and the spread across elicitations reflects decoding
variation rather than uncertainty about the user
(Section~\ref{sec::hyperparameters}).

\section{Experimental Setup}
\label{sec::exp_setup}
This section presents the datasets, the implementation details of
Agentic-Priors (elicitation settings, prior scale and language models),
and the methods we compare against.

\subsection{Datasets}
\label{sec::datasets}

We evaluate on seven publicly available choice datasets spanning
transport, vehicle purchase, consumer goods and finance
(Table~\ref{tab::dataset_characteristics}), chosen to cover different
domains and because each records user characteristics that can
serve as $z_i$; Table~\ref{tab::elicitation}
lists the quantities the LLM states for each. All are stated-choice
experiments except LPMC, which records revealed choices, i.e.\ real
trips. These choices come from surveys rather than from users of a deployed
system; for consistency with the rest of the paper, we refer to their
respondents as users. From each dataset we randomly draw 100 users as a held-out test set; all reported results are on these
users, and the HBM baselines are fitted on all remaining
(non-test) users. In each of five random permutations of a test user's tasks,
three are held out and the rest revealed one at a time; at every $T$ we
score posterior-predictive accuracy and log-loss on the held-out tasks.

\begin{table*}[t]
\centering
\caption{Datasets. Tasks: choice tasks per user.}
\label{tab::dataset_characteristics}
\small
\begin{tabularx}{\textwidth}{l r c l X X}
  \toprule
  Dataset & Users & Tasks & Alternatives & Choice attributes & $z_i$ \\
  \midrule
  Swissmetro & 1{,}191 & 9
  & Train, Swissmetro, Car\textsuperscript{a}
  & Travel time, cost; Train and Car ASCs
  & Age, gender, income, trip purpose, travel class, luggage, who pays,
    rail season ticket \\
  LPMC & 1{,}304 & 6--16
  & Walk, cycle, public transport, drive
  & Mode-specific travel times, cost, PT interchanges; mode ASCs
  & Age, gender, travel year, driving licence, household car ownership,
    fare type, trip purpose \\
  Cars / Cars CN & 384 / 448 & 15
  & Three vehicles
  & Powertrain (HEV/PHEV/BEV vs gasoline), electric range, fast
    charging, price, operating cost, acceleration, brand origin
  & Age, gender, household income, marital status, education,
    household size, vehicles owned, daily mileage, home/work charging
    access, environmental and status attitudes; urbanicity (US) or
    city (China) \\
  EV & 983 & 6
  & Gasoline car, EV
  & EV indicator, price, range, walking time to home/work charger,
    fast-charge time, in-town fast-charger drive time and availability,
    highway charger spacing and availability
  & Age, gender, state, household income, education, car budget,
    daily mileage, home/work charging access, EV ownership \\
  Beer & 1{,}291 & 8
  & Two six-packs, no purchase
  & Packaging, water/carbon/landfill eco-labels, locally brewed, price;
    no-purchase ASC
  & Age, region, income, education, primary shopper, purchase frequency,
    typical price, usual beer type, novelty seeking, recycling \\
  Bank & 946 & 14--17
  & Two credit-card offers
  & Interest rate, reward programme, annual fee, issuing bank, rebate,
    credit line, grace period (levels disguised)
  & Age, gender, income \\
  \bottomrule
\end{tabularx}
\\[2pt]{\footnotesize \textsuperscript{a}Car available only to car
owners.}
\end{table*}

\begin{table}[t]
\centering
\caption{LLM-stated quantities and their reference scale.}
\label{tab::elicitation}
\small
\begin{tabularx}{\columnwidth}{l l X}
  \toprule
  Dataset & Reference & Quantities stated \\
  \midrule
  Swissmetro & Cost (CHF)
  & Value of time; Train and Car vs Swissmetro preference \\
  LPMC & Cost (\pounds)
  & Cycle, PT and drive vs walk; value of time per mode; avoiding an
    interchange \\
  Cars & Price (\$ / \textyen)
  & Each powertrain vs gasoline; per unit of electric range; fast
    charging; lower operating cost; faster acceleration; each brand
    origin vs German \\
  EV & Price (\$)
  & EV vs gasoline; per mile of range; per minute of home/work walking,
    fast-charge and in-town charger time; in-town and highway charger
    availability; per mile of highway charger spacing \\
  Beer & Price (\$)
  & Bottles vs cans; each eco-label; locally brewed; highest price for
    a basic six-pack \\
  Bank & Rating ($\pm 100$)
  & Desirability of each attribute level relative to its baseline \\
  \bottomrule
\end{tabularx}
\end{table}

\subsubsection{Swissmetro}
A stated-preference survey of intercity travellers in Switzerland on a
proposed high-speed underground mag-lev (Swissmetro), chosen against
their existing train and car options~\cite{bierlaire2001acceptance}.
\subsubsection{LPMC}
Trip diaries from the London Travel Demand Survey (2012--2015), each
trip augmented with the routes, times and costs of the modes not
taken~\cite{hillel2018recreating}. We keep travellers with at least six trips.

\subsubsection{Cars and Cars CN}
A 2012--13 conjoint survey of new-car buyers in the US (Cars) and in
three Chinese cities (Cars CN), choosing among conventional, hybrid, plug-in hybrid and
battery-electric vehicles~\cite{helveston2015will}. The two markets share one
instrument, so they differ only in users (and in currency and
units, which the LLM is given in local terms).

\subsubsection{EV}
A 2019 US survey in which car buyers choose between a gasoline car and
an otherwise identical EV whose price, range and charging
infrastructure vary; prices are pivoted on each user's stated
budget~\cite{zou2020effects}.

\subsubsection{Beer}
A 2019 US survey on willingness to pay for sustainability attributes in
beer, with eco-labels and a no-purchase option~\cite{staples2020consumer}.

\subsubsection{Bank}
A credit-card conjoint in which customers choose between pairs of
offers~\cite{allenby1995using}. Attribute levels are numerically disguised in the
public data, so the LLM rates each level's desirability instead of
stating a monetary value (Section~\ref{sec::mon_eq_fmrk}).

\begin{table*}[t]
  \centering
  \caption{Cold start ($T = 0$) accuracy (\%, higher is better) and
  log-loss (lower is better) on the test users ($n = 100$ per
  dataset). Methods above the line use no historical choice data;
  ``+ $z_i$'' methods are also given the user's information.
  Best no-data method per column in bold.}
  \label{tab::cold_start}
  \small
  \begin{tabular}{l rrrrrrr}
    \toprule
    & Swissmetro & LPMC & Cars & Cars CN & EV & Beer & Bank \\
    \midrule
    \multicolumn{8}{l}{\emph{Accuracy}} \\
    Uninformed\textsuperscript{a} & 25.8 & 21.6 & 33.6 & 36.3 & 48.5 & 35.1 & 51.0 \\
    AutoElicit & \textbf{69.3} & 22.9 & 48.9 & \textbf{46.0} & 41.1 & 27.3 &
\textbf{66.1} \\
    AutoElicit + $z_i$ & 68.5 & 45.7 & 51.5 & 44.7 & 41.1 & 35.3 & 65.7 \\
    Agentic-Priors & 55.8 & 46.8 & \textbf{56.5} & 44.6 & \textbf{55.9} & 40.6 & 65.9 \\
    Agentic-Priors + $z_i$ & 62.5 & \textbf{61.9} & 52.3 & 43.6 & 53.5 & 48.2 & 62.7 \\
    In-Context Learning + $z_i$ & 56.7 & 44.1 & 47.6 & 41.6 & 49.4 & \textbf{48.3} & 56.1 \\
    \cmidrule{1-8}
    HBM & 68.5 & 71.1 & 64.0 & 50.5 & 65.8 & 46.1 & 74.1 \\
    HBM + $z_i$ & 70.3 & 78.3 & 64.1 & 50.1 & 66.0 & 54.0 & 74.5 \\
    \midrule
    \multicolumn{8}{l}{\emph{Log-loss}} \\
    Uninformed & 1.092 & 1.405 & 1.101 & 1.100 & 0.697 & 1.098 & 0.693 \\
    AutoElicit & 0.821 & 1.583 & 1.220 & 1.313 & 1.276 & 1.314 & 0.758 \\
    AutoElicit + $z_i$ & \textbf{0.800} & 1.453 & 1.125 & 1.165 & 1.310 & 1.245 & 0.762 \\
    Agentic-Priors & 0.969 & 1.294 & \textbf{0.962} & \textbf{1.077} & \textbf{0.686} & 1.085 & \textbf{0.666} \\
    Agentic-Priors + $z_i$ & 0.853 & \textbf{1.128} & 0.989 & 1.150 & 0.689 & \textbf{1.048} & 0.672 \\
    In-Context Learning + $z_i$ & 0.956 & 1.392 & 1.079 & 1.215 & 1.159 & 1.056 & 0.685 \\
    \cmidrule{1-8}
    HBM & 0.753 & 0.762 & 0.797 & 0.985 & 0.629 & 1.042 & 0.549 \\
    HBM + $z_i$ & 0.725 & 0.625 & 0.789 & 0.987 & 0.623 & 0.932 & 0.548 \\
    \bottomrule
  \end{tabular}
  \\[2pt]{\footnotesize \textsuperscript{a}Below chance on Swissmetro
  (35\%) and LPMC (25\%): averaging predictions over a zero-mean prior
  favours alternatives with extreme attribute values.}
\end{table*}

\subsection{Agentic-Priors Implementation Details}
\label{sec::hyperparameters}

\subsubsection{Elicitation settings} We set $c_{\mathrm{ref}} = 1$ for all datasets:
where a numeraire exists, its coefficient is $-1$ in normalised units;
on Bank, a maximally desirable level has raw utility $1$. Each
user is elicited with one call (the prompt of
Figure~\ref{fig::prompt}, temperature $0.4$, fixed seed). A reply
that cannot be parsed, or that implies $|\beta_{ik}^{\mathrm{norm}}| > 25$, is
reissued with a new seed and an instruction to return only the required
output, at most twice. All responses are cached, and every reported
result replays from the cache; the code and the cache will be made
publicly available.

\subsubsection{Prior scale} We set $\sigma = 1$, which we selected on a development set of 100 users from Swissmetro (different from the test set) and held fixed for all other datasets and
experiments; Swissmetro is therefore our development dataset. In
ablations on Swissmetro (Appendix~\ref{app::ablations}), the LLM's stated
ranges for each $a_{ik}$ had to be widened about $16\times$ before the
prior learned as fast as the isotropic one, and never significantly
outperformed it in log-loss. Averaging eight
elicitations per user did not help either: with the same
$\sigma$ its log-loss was worse at every $T$, and using the eight
answers' own spread as the covariance made the prior too narrow to
learn from (log-loss after six choices $0.765$ vs.\ $0.624$).

\subsubsection{Language Models}

Our primary elicitor is Qwen3-32B (4-bit AWQ, reasoning disabled), chosen
because it is open-weight, relatively small and cheap to run, so a prior
can be elicited for every new user without a proprietary API. To
test sensitivity to this choice, we repeat cold start elicitation on
three datasets covering both reference scales (Swissmetro, Cars
and Bank) with seven open-weight models from four families (Mistral,
Qwen, Gemma, Llama; 8--31B) on a public API, with identical prompts and
no per-model tuning (full results in Appendix~\ref{app::crossmodel}).
Cold start accuracy is largely insensitive to the
elicitor: on Swissmetro, every model except Gemma-3-12B is within $2$
percentage points of our primary model (none significant), and on Cars
and Bank the others differ by at most $4$ points in either direction.
Log-loss is more sensitive, since it also penalises miscalibrated
magnitudes: on Swissmetro, only Mistral-Small-24B matches our primary
model, the other 24--31B models are $0.03$--$0.06$ worse, and
Llama-3.1-8B is $0.10$ worse. The exception on both metrics is the
previous-generation Gemma-3-12B, which states implausibly large
monetary values, loses $5$--$6$ accuracy points on Swissmetro and Cars,
and is worse than an uninformed prior in log-loss on both; Gemma-4
does not show this failure.

\begin{figure*}[t]
  \centering
  \includegraphics[width=\textwidth]{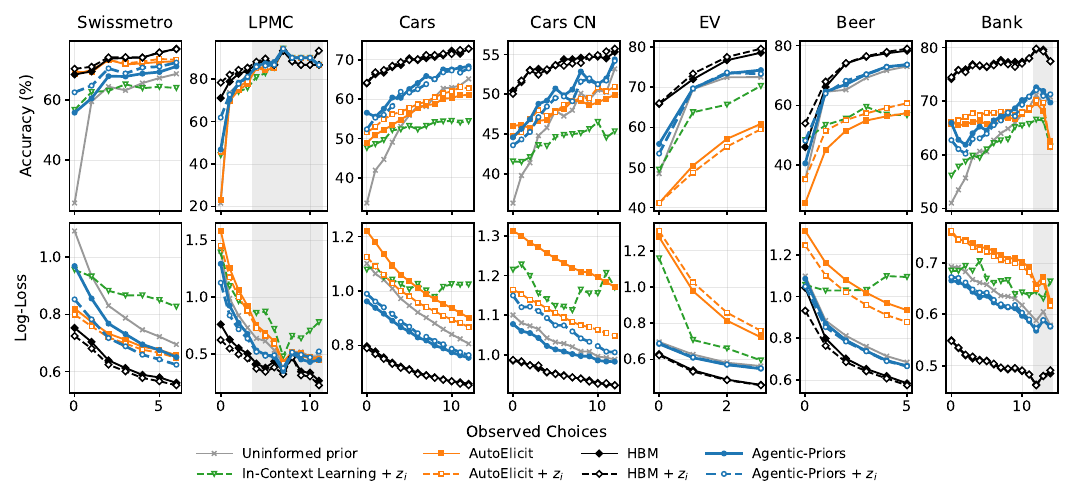}
  \caption{Accuracy and log-loss as choices are observed.
  Dashed lines with open markers: methods given the user's
  information $z_i$. Shaded: steps reached by only some users,
  whose points average a smaller, self-selected group.}
  \label{fig::learning}
\Description{Fourteen line plots in two rows, accuracy above and log-loss
below, one column per dataset, showing each method's performance as the
number of observed choices grows from zero. The HBM curves are highest
in accuracy and lowest in log-loss almost throughout; Agentic-Priors
lies between them and the uninformed prior at the start, and in-context
learning falls further behind on four datasets as choices are observed.}
\end{figure*}

\begin{figure*}[t]
  \centering
  \includegraphics[width=\textwidth]{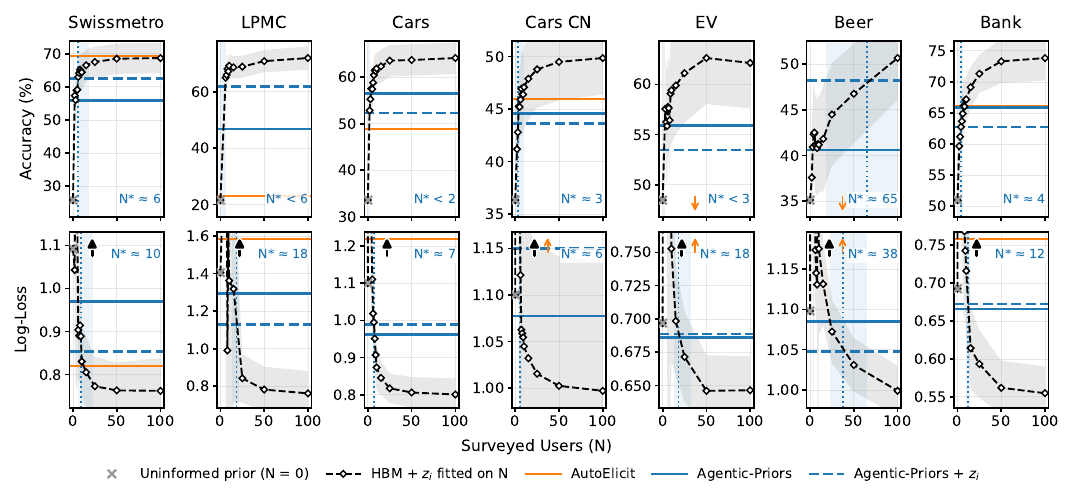}
  \caption{Cold start performance of HBM + $z_i$ fitted to $N$ surveyed
  users ($N = 0$: uninformed prior; grey band: 95\% interval).
  $N^*$: survey size at which the HBM overtakes Agentic-Priors + $z_i$
  (blue band: bootstrap interval; $N^* < n$: already ahead at the smallest
  survey that can be fitted). Arrows: values beyond the plotted range.}
  \label{fig::exchange}
\Description{Fourteen line plots in two rows, accuracy above and log-loss
below, one column per dataset. Each shows the cold start performance of
an HBM fitted to a growing number of surveyed users, from zero to
one hundred, with horizontal lines for Agentic-Priors with and without
user information and for AutoElicit. In log-loss, HBMs fitted on
the smallest surveys lie far above the plotted range, marked by a black
arrow at the top edge. The HBM curve overtakes the Agentic-Priors line
within about 40 users on every dataset except Beer in accuracy
(about 65).}
\end{figure*}

\subsection{Benchmarks}
\label{sec::benchmarks}

We compare against six benchmarks with varying data requirements
(Table~\ref{tab::prior_information}). To separate what the LLM knows
about the task from what it infers about the individual, we also run
Agentic-Priors without $z_i$, where the LLM sees only the task
description. All methods
except In-Context Learning supply a Gaussian prior to the same
Bayesian learner (Section~\ref{sec::bayes_update}) and are evaluated
identically.

\begin{table}[t]
\centering
\caption{Information required by each method.}
\label{tab::prior_information}
\small
\begin{tabular}{lcc}
  \toprule
  Method & Historical choices & $z_i$ \\
  \midrule
  Uninformed prior & No & No \\
  AutoElicit & No & No \\
  AutoElicit + $z_i$ & No & Yes \\
  In-Context Learning + $z_i$ & No & Yes \\
  HBM & Yes & No \\
  HBM + $z_i$ & Yes & Yes \\
  \midrule
  Agentic-Priors (ours) & No & No \\
  Agentic-Priors + $z_i$ (ours) & No & Yes \\
  \bottomrule
\end{tabular}
\end{table}

\subsubsection{Uninformed Prior}

Without historical data, one can resort to an uninformed prior,
$\beta_i \sim \mathcal{N}(0, \sigma^2 I)$
(Section~\ref{sec::conventional_priors}). We set $\sigma = 1$, the
natural unit scale under design-SD normalisation, so it differs from
our prior only in its mean.

\subsubsection{Hierarchical Bayes Model}

We fit the hierarchical model of Section~\ref{sec::conventional_priors}
on the non-test users' choices with the hierarchical MNL
sampler of the \texttt{bayesm} R package~\cite{rossi2003bayesian} (a
single normal mixing component, default hyperpriors; 6{,}000 MCMC
draws, 1{,}000 burn-in), and use the posterior means of
$\mu^{\mathrm{pop}}$ and $\Sigma^{\mathrm{pop}}$ as the prior for every test
user. HBM + $z_i$ conditions the mean on the user facts
shown to the LLM,
$\beta_i \sim \mathcal{N}(\mu^{\mathrm{pop}} + \Delta^T z_i, \Sigma^{\mathrm{pop}})$,
with $z_i$ standardised and centred on the non-test users; to
keep $\Delta$ from overfitting, we place a $\mathcal{N}(0, 0.01\,I)$
prior on $\operatorname{vec}(\Delta)$. The covariance stays shared
across the population. For the exchange rate
(Section~\ref{sec::res_exchange}), we also fit HBM + $z_i$ to random,
nested surveys of $N = 2$ to 400 non-test users (10 draws per $N$
up to 50, 3 above; Appendix~\ref{app::exchange}).

\subsubsection{AutoElicit}

AutoElicit~\cite{capstick2025autoelicit} is the closest existing
LLM-based prior elicitation method. We run the authors' code, adapting
only the prompt template: AutoElicit targets linear and logistic
regression, so the template for choice tasks is ours, with role
paraphrases generated by the authors' rephrasing prompt. The same LLM
as for our method is prompted with every combination of $10$
paraphrased system roles and $10$ user roles ($S = 100$ elicitations,
temperature $0.1$), each time stating a Normal mean and standard
deviation for each parameter. Following the authors, the prior is the
diagonal Gaussian that moment-matches the equal-weight mixture of the
$S$ elicited Normals, common to all users. As in their
implementation, intercepts are not elicited but receive a fixed
$\mathcal{N}(0, 1)$ prior on the unnormalised dummy scale. AutoElicit +
$z_i$ adds $z_i$ to every user-role prompt and moment-matches per
user; since the intercepts keep their fixed prior, $z_i$ affects
only the attribute coefficients.

\subsubsection{In-Context Learning + $z_i$}

To test whether a prior and a choice model are needed at all, the same
LLM predicts choices directly. For user $i$, the prompt contains
$z_i$, the $T$ choices observed so far (the same ones the Bayesian
learner sees) and the held-out tasks, and the LLM returns a probability
for each alternative: learning happens only in context. We use our
method's system prompt and decoding settings (one call per user
and value of $T$), score the same held-out tasks, and mix $10^{-3}$ of
uniform probability into each prediction so that a confidently wrong
answer has finite log-loss.

\section{Results}
\label{sec::results}

We address three questions. First, at cold start ($T = 0$), is a prior
elicited by Agentic-Priors more accurate than the alternatives that
need no historical data (Section~\ref{sec::res_cold})? Second, does its
advantage persist as the user's choices are observed, and is updating
an explicit prior better than giving the LLM those choices in context
(Section~\ref{sec::res_learning})? Third, how much population data is
the prior worth (Section~\ref{sec::res_exchange})? Unless otherwise stated, all comparisons are paired over the same 100 test users per dataset, and ``significant'' means $p < 0.05$ in a paired $t$-test, Holm-corrected within dataset (Appendix~\ref{app::paired}).

\subsection{Cold Start}
\label{sec::res_cold}

Table~\ref{tab::cold_start} shows results at $T = 0$. Agentic-Priors + $z_i$
is significantly more accurate than the uninformed prior on all datasets, by 5 to 40 percentage
points. In accuracy, it beats AutoElicit on LPMC, EV and Beer and loses
only on Swissmetro; adding $z_i$ to AutoElicit changes no accuracy
verdict in its favour. In-Context Learning is less accurate on six
datasets, significantly on LPMC and Bank, and never significantly better. Both HBMs remain
ahead, except on Beer, where Agentic-Priors + $z_i$ is level with the
HBM without $z_i$ (no significant difference) using no data.
Among methods that need no historical data, Agentic-Priors + $z_i$ is
thus the most consistent at cold start: always more accurate than the
uninformed prior, never significantly behind In-Context Learning, and
behind AutoElicit only on Swissmetro.
Swapping covariances between Agentic-Priors + $z_i$ and the HBM shows
that its remaining gap to the HBM at cold start lies in the prior's mean: with the HBM's
covariance, our prior's accuracy changes by at most 2 points on every
dataset. The LLM's stated trade-offs, not the spread
placed around them, are what population data improves on
(Appendix~\ref{app::diagnostics}).

\subsection{Learning From Observed Choices}
\label{sec::res_learning}

Figure~\ref{fig::learning} shows performance as the user's choices are
observed. The accuracy advantage over the uninformed prior shrinks
quickly: on LPMC, EV and Beer a single choice is enough to close it
(in log-loss it persists), on Cars CN it falls to 1--3 points, while on
Swissmetro, Cars and Bank it persists (2 to 4 percentage points at the
last step all users reach). In-Context Learning falls further
behind on Swissmetro, Cars, Cars CN and Beer, by up to 17 points; on
the others it improves about as fast as the Bayesian learner, but from
a lower start. The HBMs stay ahead (on Beer, from the first observed
choice), though the accuracy gap narrows as the user's own choices
outweigh the population prior.
The prior's advantage thus persists in accuracy on three datasets and
fades on the others, and updating it with the user's choices is never
significantly worse than giving them to the LLM in context.
What slows Agentic-Priors after
cold start, compared with the HBMs, is its isotropic covariance, not the LLM's estimates: with
the HBM's covariance, our mean learns faster on every dataset but
Swissmetro, and after the last observed choice it is within about 1
point of the HBM's accuracy on LPMC, Cars CN, Beer and Bank
(Appendix~\ref{app::diagnostics}).

\subsection{Exchange Rate with Population Data}
\label{sec::res_exchange}

Figure~\ref{fig::exchange} compares Agentic-Priors + $z_i$ with HBMs + $z_i$
fitted to $N$ surveyed users; the survey size $N^*$ at which the
HBM overtakes the prior is its exchange rate (all values in
Appendix~\ref{app::exchange}). In log-loss, the prior is
worth about 10 users on Swissmetro, 7 on Cars and 12--38 on the
others (Beer highest); on Cars CN, where it is worse than the uninformed
prior, it beats only HBMs fitted on fewer than about 6 users,
whose priors are badly over-confident. In accuracy, it is worth about 6
users on Swissmetro, 4 on Bank and 3 on Cars CN, and fewer than
the smallest survey we could fit on Cars (2), EV (3) and LPMC (6); only
on Beer is it worth more (about 65). In log-loss, the prior without
$z_i$ is worth about as much on every dataset except Swissmetro (6) and
Beer (22), and the HBM gains little from $z_i$ either, outside
LPMC and Beer: the user information these datasets record changes
little in either method's exchange rate.
The prior is therefore worth a small survey: at most about 40
users in log-loss, and at most about 6 in accuracy on all datasets
but Beer.
It is this small
because a small survey quickly estimates the mean that the LLM must
guess: our mean combined with the covariance of an HBM without $z_i$
fitted on $N$ users beats that HBM in log-loss only for the
smallest surveys (EV at $N = 10$, Beer up to $N = 15$), and from 25
users the survey's own mean is as good or better on every dataset
(Appendix~\ref{app::diagnostics}).

\section{Conclusions and Future Work}
\label{sec::conclusion}

We introduced Agentic-Priors, in which an LLM states a user's trade-offs on a reference scale, such as
willingness to pay, that we map to a prior over choice-model
coefficients. Across seven datasets, the prior is more accurate than an
uninformed prior at cold start on every dataset, beats AutoElicit in accuracy on
three, and outperforms in-context learning with the same information.
Its value is real but bounded: in log-loss it is worth between about 6
and 40 surveyed users, and most of it comes from
the LLM's knowledge of the task rather than of the individual. An LLM
prior is therefore best seen as a way to start learning before any data
exist, not a substitute for population data once a small survey is
available.

Future work should extend the LLM's role beyond the prior: an
LLM-assisted learner could choose which choices to ask about or
revise the prior when observed choices contradict it, while keeping
the Bayesian update that in-context learning lacks. The prior's
covariance is also set without input from the LLM; since its stated uncertainty
proved far too narrow, a dedicated protocol for eliciting calibrated
spread and correlations is a natural next step.


\begin{acks}
This work was supported by a Turing AI Fellowship (EP/V022067/1) on
Citizen-Centric AI Systems (\url{https://ccais.ac.uk/}) and by the FEVER
Programme Grant (EP/W005883/1).
\end{acks}



\bibliographystyle{ACM-Reference-Format}
\bibliography{refs}

\clearpage
\onecolumn
\raggedbottom
\appendix
\tcbset{breakable}

\section{Prompts}
\label{app::prompts}

This appendix lists, for every dataset, the exact messages Agentic-Priors sends to the
LLM. As in the main text, we refer to each dataset's survey respondents as users.
Each user is elicited with one system message and one user message, at
temperature $0.4$ with a fixed seed. The system message is fixed per dataset. The user
message gives the user's profile $z_i$ as a sentence, states that no past choices
are available (all reported results elicit the prior at $T = 0$), and lists the
quantities to state on the dataset's reference scale. We show the user message for the
first test user of each dataset; other users differ only in the profile
sentence. When $z_i$ is omitted, the profile sentence is replaced by the generic line
shown last. The texts below are generated from the code that sends the requests and
match the logged requests exactly.

{\let\clearpage\relax
\paragraph{Retry instruction.} If a reply cannot be parsed, or implies an implausibly large coefficient, the request is reissued (at most twice) with this text appended to the system message:
\begin{rolebox}{System (appended on retries)}{sysbg}{sysaccent}
\begin{promptverb}
STRICT OUTPUT: reply with exactly one complete, valid JSON object — double-quoted keys, plain numbers (no '+' signs), no comments, no trailing commas, nothing after the closing brace.
\end{promptverb}
\end{rolebox}

\clearpage
\subsection{Swissmetro}
\label{app::prompt_swissmetro}

\begin{promptbox}{LLM input \textperiodcentered\ Swissmetro}
\begin{rolebox}{System}{sysbg}{sysaccent}
\begin{promptverb}
You are a transport-choice analyst estimating one traveller's preferences for a discrete-choice model of Train vs Swissmetro (a fast mag-lev) vs Car. Travel time and cost reduce utility. Reply with ONLY a JSON object, no prose.
\end{promptverb}
\end{rolebox}
\vspace{-1mm}
\begin{rolebox}{User message (one test user, $T = 0$)}{userbg}{useraccent}
\begin{promptverb}
Traveller profile: a man aged 39-54, annual income over CHF 100,000, travelling for business travel, who normally travels first class, with no luggage, sharing the fare with their employer.

No past choices are available for this traveller.

Estimate this traveller's preferences as JSON with these numeric fields (money is the unit throughout):
  "value_of_time_chf_per_hour": CHF they would pay to save one hour of travel time (positive),
  "train_vs_swissmetro_chf": the extra CHF they would pay for Train over Swissmetro IF the two had identical travel time and cost — i.e. their intrinsic mode preference only, NOT the real time/cost difference (negative if they would still prefer Swissmetro),
  "car_vs_swissmetro_chf": the same for Car vs Swissmetro, again at identical travel time and cost (negative if they prefer Swissmetro).
\end{promptverb}
\end{rolebox}
\vspace{-1mm}
\begin{rolebox}{User message, profile line without $z_i$}{userbg}{useraccent}
\begin{promptverb}
Traveller profile: a Swiss intercity traveller.
\end{promptverb}
\end{rolebox}
\end{promptbox}

\clearpage
\subsection{LPMC}
\label{app::prompt_lpmc}

\begin{promptbox}{LLM input \textperiodcentered\ LPMC}
\begin{rolebox}{System}{sysbg}{sysaccent}
\begin{promptverb}
You are a transport-choice analyst estimating one London traveller's preferences for a discrete-choice model of everyday trip mode: walking vs cycling vs public transport (bus/rail/underground) vs travelling by car (as driver or passenger). Travel time and out-of-pocket cost reduce utility. Reply with ONLY a JSON object, no prose.
\end{promptverb}
\end{rolebox}
\vspace{-1mm}
\begin{rolebox}{User message (one test user, $T = 0$)}{userbg}{useraccent}
\begin{promptverb}
Traveller profile: a 63-year-old man living in London, making everyday trips in 2014, holding a driving licence, in a household with at least one car per adult, holding a 60+/Freedom Pass giving free public transport, mostly everyday errands and other personal trips.

No past trips are available for this traveller.

Estimate this traveller's preferences as JSON with these numeric fields:
All amounts are GBP (British pounds): the most the traveller would pay (negative = they would need to be paid that much instead). Fields:
  "cycle_vs_walk_gbp": £ to cycle rather than walk, for a trip that takes 20 minutes door to door either way and costs the same,
  "pt_vs_walk_gbp": £ to take public transport (a direct service with no interchanges) rather than walk, for a trip that takes 20 minutes door to door either way and costs the same,
  "car_vs_walk_gbp": £ to travel by car, as driver or passenger, rather than walk, for a trip that takes 20 minutes door to door either way and costs the same,
  "vot_walk_gbp_per_hour": £ to avoid one hour of walking,
  "vot_cycle_gbp_per_hour": £ to avoid one hour of cycling,
  "vot_pt_gbp_per_hour": £ to avoid one hour on public transport door to door,
  "vot_car_gbp_per_hour": £ to avoid one hour of travelling by car, as driver or passenger,
  "interchange_gbp": £ for one additional public-transport interchange on the same trip, with the same door-to-door time and cost.
\end{promptverb}
\end{rolebox}
\vspace{-1mm}
\begin{rolebox}{User message, profile line without $z_i$}{userbg}{useraccent}
\begin{promptverb}
Traveller profile: a London traveller making everyday trips.
\end{promptverb}
\end{rolebox}
\end{promptbox}

\clearpage
\subsection{Cars}
\label{app::prompt_cars}

\begin{promptbox}{LLM input \textperiodcentered\ Cars}
\begin{rolebox}{System}{sysbg}{sysaccent}
\begin{promptverb}
You are a vehicle-choice analyst estimating one US car buyer's preferences for a discrete-choice model of new cars that vary in powertrain (conventional gasoline, hybrid, plug-in hybrid, or battery-electric), electric range, fast-charging, price, operating cost, 0-60 acceleration, and brand country-of-origin. Price and operating cost reduce utility. Reply with ONLY a JSON object, no prose.
\end{promptverb}
\end{rolebox}
\vspace{-1mm}
\begin{rolebox}{User message (one test user, $T = 0$)}{userbg}{useraccent}
\begin{promptverb}
Buyer profile: a 45-year-old man, with a household income of about $81,000, married, a college graduate, living in a rural area, in a 2-person household with no children, currently owning 3 vehicles, driving about 40 miles a day, able to charge an electric car at home, who sees a car's appearance as an important status symbol.

No past choices are available for this buyer.

Estimate this buyer's car preferences as JSON with these numeric fields:
All amounts are US dollars of purchase price (positive = would pay more; negative = would need a discount). Fields:
  "hev_vs_gasoline_usd": extra $ for a hybrid (HEV) vs an otherwise-identical conventional gasoline car,
  "phev_vs_gasoline_usd": same for a plug-in hybrid at its base 10-mile electric range,
  "bev_vs_gasoline_usd": same for a battery-electric car at its base 75-mile range,
  "phev_range_usd_per_mile": extra $ per additional mile of plug-in-hybrid electric range,
  "bev_range_usd_per_mile": extra $ per additional mile of battery-electric range,
  "phev_fastcharge_usd": extra $ for fast-charging capability on a plug-in hybrid,
  "bev_fastcharge_usd": extra $ for fast-charging capability on a battery-electric car,
  "opcost_usd_per_cent_per_mile": extra $ of purchase price to lower operating cost by 1 cent per mile,
  "acceleration_usd_per_second": extra $ for each 1-second-quicker 0-60 mph time,
  "american_vs_german_usd": extra $ for an American-brand vs a German-brand car,
  "japanese_vs_german_usd": same for Japanese vs German,
  "chinese_vs_german_usd": same for Chinese vs German,
  "skorean_vs_german_usd": same for South-Korean vs German.
\end{promptverb}
\end{rolebox}
\vspace{-1mm}
\begin{rolebox}{User message, profile line without $z_i$}{userbg}{useraccent}
\begin{promptverb}
Buyer profile: a US car buyer.
\end{promptverb}
\end{rolebox}
\end{promptbox}

\clearpage
\subsection{Cars CN}
\label{app::prompt_cars_china}

\begin{promptbox}{LLM input \textperiodcentered\ Cars CN}
\begin{rolebox}{System}{sysbg}{sysaccent}
\begin{promptverb}
You are a vehicle-choice analyst estimating one Chinese car buyer's preferences for a discrete-choice model of new cars sold in urban China that vary in powertrain (conventional gasoline, hybrid, plug-in hybrid, or battery-electric), electric range, fast-charging, price, running cost, acceleration, and brand country-of-origin. Price and running cost reduce utility. Reply with ONLY a JSON object, no prose.
\end{promptverb}
\end{rolebox}
\vspace{-1mm}
\begin{rolebox}{User message (one test user, $T = 0$)}{userbg}{useraccent}
\begin{promptverb}
Buyer profile: a 47-year-old woman in Shanghai, with an annual household income of about ¥120,000, married, without a college degree, in a 3-person household with 1 child, currently owning 1 vehicle, driving about 35 km a day, with no access to electric-car charging at home or work.

No past choices are available for this buyer.

Estimate this buyer's car preferences as JSON with these numeric fields:
All amounts are Chinese yuan (¥, RMB) of purchase price (positive = would pay more; negative = would need a discount). Fields:
  "hev_vs_gasoline_rmb": extra ¥ for a hybrid (HEV) vs an otherwise-identical conventional gasoline car,
  "phev_vs_gasoline_rmb": same for a plug-in hybrid at its base 16-km electric range,
  "bev_vs_gasoline_rmb": same for a battery-electric car at its base 121-km range,
  "phev_range_rmb_per_km": extra ¥ per additional km of plug-in-hybrid electric range,
  "bev_range_rmb_per_km": extra ¥ per additional km of battery-electric range,
  "phev_fastcharge_rmb": extra ¥ for fast-charging capability on a plug-in hybrid,
  "bev_fastcharge_rmb": extra ¥ for fast-charging capability on a battery-electric car,
  "opcost_rmb_per_rmb_100km": extra ¥ of purchase price to lower running cost by ¥1 per 100 km,
  "acceleration_rmb_per_second": extra ¥ for each 1-second-quicker 0-100 km/h time,
  "american_vs_german_rmb": extra ¥ for an American-brand vs a German-brand car,
  "japanese_vs_german_rmb": same for Japanese vs German,
  "chinese_vs_german_rmb": same for Chinese vs German,
  "skorean_vs_german_rmb": same for South-Korean vs German.
\end{promptverb}
\end{rolebox}
\vspace{-1mm}
\begin{rolebox}{User message, profile line without $z_i$}{userbg}{useraccent}
\begin{promptverb}
Buyer profile: a car buyer in urban China.
\end{promptverb}
\end{rolebox}
\end{promptbox}

\clearpage
\subsection{EV}
\label{app::prompt_ev2019}

\begin{promptbox}{LLM input \textperiodcentered\ EV}
\begin{rolebox}{System}{sysbg}{sysaccent}
\begin{promptverb}
You are a vehicle-choice analyst estimating one US car buyer's preferences in 2019 for a discrete-choice model of choosing between a conventional gasoline car and an otherwise-identical electric vehicle that differ in price, driving range, and the local charging environment (walking distance to slow charging, fast-charge time, and fast-charger availability in town and on highways). Price reduces utility. Reply with ONLY a JSON object, no prose.
\end{promptverb}
\end{rolebox}
\vspace{-1mm}
\begin{rolebox}{User message (one test user, $T = 0$)}{userbg}{useraccent}
\begin{promptverb}
Buyer profile: a 25-year-old man in Florida, shopping for a car in 2019, with a household income $25,000-$49,999, with some college, with a car budget of about $10,000, driving about 5 miles a day, without charging access at their home parking.

No past choices are available for this buyer.

Estimate this buyer's preferences as JSON with these numeric fields:
All amounts are US dollars of purchase price (positive = would pay more; negative = would need a discount). Fields:
  "ev_vs_gasoline_usd": extra $ for an electric vehicle vs an otherwise-identical gasoline car, at equal price, 300-mile range and convenient charging,
  "range_usd_per_mile": extra $ per additional mile of EV driving range,
  "homewalk_usd_per_minute": extra $ to have home slow charging one walking-minute closer,
  "workwalk_usd_per_minute": same for charging at work,
  "fastcharge_usd_per_minute": extra $ for a one-minute-faster empty-to-full fast charge,
  "town_usd_per_minute": extra $ to have the nearest in-town fast charger one driving-minute closer,
  "town_available_usd": extra $ for having in-town fast charging at all vs none,
  "highway_usd_per_mile": extra $ to reduce the distance between highway fast chargers by one mile,
  "highway_available_usd": extra $ for having highway fast charging at all vs none.
\end{promptverb}
\end{rolebox}
\vspace{-1mm}
\begin{rolebox}{User message, profile line without $z_i$}{userbg}{useraccent}
\begin{promptverb}
Buyer profile: a US car buyer in 2019.
\end{promptverb}
\end{rolebox}
\end{promptbox}

\clearpage
\subsection{Beer}
\label{app::prompt_beer2019}

\begin{promptbox}{LLM input \textperiodcentered\ Beer}
\begin{rolebox}{System}{sysbg}{sysaccent}
\begin{promptverb}
You are a consumer-choice analyst estimating one US beer shopper's preferences for a discrete-choice model of six-pack purchases. Six-packs differ in price, packaging (glass bottles vs aluminum cans), three sustainability eco-labels (water conservation, carbon reduction, landfill diversion), and whether the beer is locally brewed; the shopper may also buy nothing. Higher price reduces utility. Reply with ONLY a JSON object, no prose.
\end{promptverb}
\end{rolebox}
\vspace{-1mm}
\begin{rolebox}{User message (one test user, $T = 0$)}{userbg}{useraccent}
\begin{promptverb}
Shopper profile: a US beer shopper aged 25-34, living in the US West, with a household income $50,000-$74,999, with some college but no degree, the household's primary grocery shopper, who never buys beer, regularly recycles at home.

No past choices are available for this shopper.

Estimate this shopper's preferences as JSON with these numeric fields:
All amounts are US dollars per six-pack (positive = would pay more; negative = would need a discount). Fields:
  "bottles_vs_cans_usd": extra $ for a six-pack in glass bottles instead of aluminum cans,
  "water_label_usd": extra $ for an eco-label certifying reduced water use and wastewater in brewing,
  "carbon_label_usd": extra $ for an eco-label certifying reduced energy use and carbon emissions,
  "landfill_label_usd": extra $ for an eco-label certifying landfill diversion and reduced solid waste,
  "local_usd": extra $ for a beer that is locally brewed,
  "max_basic_sixpack_usd": the highest price in $ at which this shopper would still buy a basic six-pack (aluminum cans, no eco-labels, not locally brewed) rather than buy no beer.
\end{promptverb}
\end{rolebox}
\vspace{-1mm}
\begin{rolebox}{User message, profile line without $z_i$}{userbg}{useraccent}
\begin{promptverb}
Shopper profile: a US beer shopper in 2019.
\end{promptverb}
\end{rolebox}
\end{promptbox}

\clearpage
\subsection{Bank}
\label{app::prompt_bank}

\begin{promptbox}{LLM input \textperiodcentered\ Bank}
\begin{rolebox}{System}{sysbg}{sysaccent}
\begin{promptverb}
You are a financial-choice analyst estimating one bank customer's preferences for a discrete-choice model of credit-card offers that differ in interest rate, reward program, annual fee, issuing bank, rebate, credit line, and grace period. A card is generally more attractive with a lower interest rate, a lower annual fee, a bigger rebate, a higher credit line, and a longer grace period. Reply with ONLY a JSON object, no prose.
\end{promptverb}
\end{rolebox}
\vspace{-1mm}
\begin{rolebox}{User message (one test user, $T = 0$)}{userbg}{useraccent}
\begin{promptverb}
Customer profile: a 50-year-old man, with an annual income of about $40,000.

No past choices are available for this customer.

Estimate this customer's credit-card preferences as JSON with these fields:
Rate EACH field on a scale from -100 to +100: +100 = this level makes the card MUCH more attractive to this customer than the reference level, -100 = MUCH less attractive, 0 = no difference. Note on the reward programs: each is a CONDITIONAL benefit (an annual-fee waiver or interest-rebate reduction that the customer earns only by meeting a required level of card spending or checking-account balance), so more is not automatically better. Fields:
  "medium_fixed_interest": desirability of a MEDIUM fixed interest rate instead of a HIGH fixed rate,
  "low_fixed_interest": desirability of a LOW fixed interest rate instead of a HIGH fixed rate,
  "medium_variable_interest": desirability of a MEDIUM VARIABLE interest rate (which can change over time) instead of a HIGH fixed rate,
  "reward_program_2": desirability of conditional reward program 2 instead of reward program 1,
  "reward_program_3": desirability of conditional reward program 3 instead of reward program 1,
  "reward_program_4": desirability of conditional reward program 4 instead of reward program 1,
  "medium_annual_fee": desirability of a MEDIUM annual fee instead of a HIGH annual fee,
  "low_annual_fee": desirability of a LOW annual fee instead of a HIGH annual fee,
  "bank_b": desirability of the card issued by Bank B instead of Bank A (two comparable banks),
  "out_of_state_bank": desirability of the card issued by an unfamiliar out-of-region bank instead of a local bank,
  "medium_rebate": desirability of a MEDIUM rebate instead of a LOW rebate,
  "high_rebate": desirability of a HIGH rebate instead of a LOW rebate,
  "high_credit_line": desirability of a HIGH credit line instead of a LOW credit line,
  "long_grace_period": desirability of a LONG grace period instead of a SHORT one.
\end{promptverb}
\end{rolebox}
\vspace{-1mm}
\begin{rolebox}{User message, profile line without $z_i$}{userbg}{useraccent}
\begin{promptverb}
Customer profile: a bank customer.
\end{promptverb}
\end{rolebox}
\end{promptbox}
}

\FloatBarrier
\section{Paired Comparisons at Cold Start}
\label{app::paired}

Table~\ref{tab::cold_start} reports each method's accuracy and log-loss at $T = 0$. Tables
\ref{tab::app_paired_acc} and~\ref{tab::app_paired_ll} give the significance behind the
comparisons we draw from it: for every method, the paired difference from
Agentic-Priors + $z_i$ on the same test users, with paired $t$-test $p$-values
Holm-corrected over the 14 tests of each dataset. Each user's score is first averaged over
their five task permutations, so the unit of analysis is the user.

Agentic-Priors + $z_i$ is significantly more accurate than the uninformed prior on every dataset.
Against AutoElicit it is significantly more accurate on LPMC, EV and Beer and less accurate on
Swissmetro; with $z_i$ added to AutoElicit, the Swissmetro difference is no longer significant.
In-Context Learning + $z_i$ is less accurate on six datasets, significantly on LPMC and Bank, and
never significantly better. The HBMs are significantly more accurate on most datasets; on Beer,
the HBM without $z_i$ does not differ significantly from Agentic-Priors + $z_i$ in either metric.

\begin{table}[h]
\centering\small
\caption{Cold start ($T = 0$) paired differences from Agentic-Priors + $z_i$: Accuracy (percentage points). Method minus Agentic-Priors + $z_i$, test users ($n = 100$ per dataset); in brackets, paired $t$-test $p$-values Holm-corrected over the 14 tests of each dataset (7 methods $\times$ 2 metrics); $^*$: $p < 0.05$.}
\label{tab::app_paired_acc}
\setlength{\tabcolsep}{3pt}
\begin{tabular}{lrrrrrrr}
\toprule
  & Swissmetro & LPMC & Cars & Cars CN & EV & Beer & Bank \\
\midrule
  Uninformed & $-36.7^*$ & $-40.3^*$ & $-18.7^*$ & $-7.3^*$ & $-5.0^*$ & $-13.1^*$ & $-11.7^*$ \\
  & {\scriptsize($<$0.001)} & {\scriptsize($<$0.001)} & {\scriptsize($<$0.001)} & {\scriptsize(0.005)} & {\scriptsize(0.018)} & {\scriptsize($<$0.001)} & {\scriptsize($<$0.001)} \\[2pt]
  AutoElicit & $+6.8^*$ & $-38.9^*$ & $-3.4$ & $+2.4$ & $-12.3^*$ & $-20.9^*$ & $+3.3$ \\
  & {\scriptsize(0.049)} & {\scriptsize($<$0.001)} & {\scriptsize(0.324)} & {\scriptsize(1.000)} & {\scriptsize(0.009)} & {\scriptsize($<$0.001)} & {\scriptsize(0.091)} \\[2pt]
  AutoElicit + $z_i$ & $+5.9$ & $-16.1^*$ & $-0.8$ & $+1.1$ & $-12.3^*$ & $-12.9^*$ & $+2.9$ \\
  & {\scriptsize(0.080)} & {\scriptsize($<$0.001)} & {\scriptsize(0.723)} & {\scriptsize(1.000)} & {\scriptsize(0.009)} & {\scriptsize($<$0.001)} & {\scriptsize(0.091)} \\[2pt]
  Agentic-Priors & $-6.7^*$ & $-15.1^*$ & $+4.2$ & $+1.0$ & $+2.4$ & $-7.6^*$ & $+3.2^*$ \\
  & {\scriptsize(0.014)} & {\scriptsize($<$0.001)} & {\scriptsize(0.084)} & {\scriptsize(1.000)} & {\scriptsize(0.066)} & {\scriptsize(0.022)} & {\scriptsize($<$0.001)} \\[2pt]
  In-Context Learning + $z_i$ & $-5.8$ & $-17.7^*$ & $-4.7$ & $-2.0$ & $-4.0$ & $+0.1$ & $-6.6^*$ \\
  & {\scriptsize(0.096)} & {\scriptsize($<$0.001)} & {\scriptsize(0.084)} & {\scriptsize(1.000)} & {\scriptsize(0.181)} & {\scriptsize(1.000)} & {\scriptsize(0.002)} \\[2pt]
  HBM & $+5.9$ & $+9.2^*$ & $+11.7^*$ & $+6.9^*$ & $+12.3^*$ & $-2.1$ & $+11.4^*$ \\
  & {\scriptsize(0.080)} & {\scriptsize($<$0.001)} & {\scriptsize($<$0.001)} & {\scriptsize(0.009)} & {\scriptsize($<$0.001)} & {\scriptsize(1.000)} & {\scriptsize($<$0.001)} \\[2pt]
  HBM + $z_i$ & $+7.8^*$ & $+16.5^*$ & $+11.8^*$ & $+6.5^*$ & $+12.5^*$ & $+5.8$ & $+11.7^*$ \\
  & {\scriptsize(0.021)} & {\scriptsize($<$0.001)} & {\scriptsize($<$0.001)} & {\scriptsize(0.022)} & {\scriptsize($<$0.001)} & {\scriptsize(0.090)} & {\scriptsize($<$0.001)} \\[2pt]
\bottomrule
\end{tabular}
\end{table}

\begin{table}[h]
\centering\small
\caption{Cold start ($T = 0$) paired differences from Agentic-Priors + $z_i$: Log-loss (positive: worse than Agentic-Priors + $z_i$). Method minus Agentic-Priors + $z_i$, test users ($n = 100$ per dataset); in brackets, paired $t$-test $p$-values Holm-corrected over the 14 tests of each dataset (7 methods $\times$ 2 metrics); $^*$: $p < 0.05$.}
\label{tab::app_paired_ll}
\setlength{\tabcolsep}{3pt}
\begin{tabular}{lrrrrrrr}
\toprule
  & Swissmetro & LPMC & Cars & Cars CN & EV & Beer & Bank \\
\midrule
  Uninformed & $+0.239^*$ & $+0.278^*$ & $+0.111^*$ & $-0.049$ & $+0.008^*$ & $+0.050^*$ & $+0.020^*$ \\
  & {\scriptsize($<$0.001)} & {\scriptsize(0.001)} & {\scriptsize($<$0.001)} & {\scriptsize(1.000)} & {\scriptsize(0.008)} & {\scriptsize($<$0.001)} & {\scriptsize($<$0.001)} \\[2pt]
  AutoElicit & $-0.032$ & $+0.455^*$ & $+0.231^*$ & $+0.163$ & $+0.587^*$ & $+0.266^*$ & $+0.085$ \\
  & {\scriptsize(0.103)} & {\scriptsize($<$0.001)} & {\scriptsize($<$0.001)} & {\scriptsize(0.350)} & {\scriptsize($<$0.001)} & {\scriptsize($<$0.001)} & {\scriptsize(0.072)} \\[2pt]
  AutoElicit + $z_i$ & $-0.053^*$ & $+0.325^*$ & $+0.135^*$ & $+0.015$ & $+0.621^*$ & $+0.197^*$ & $+0.090$ \\
  & {\scriptsize(0.049)} & {\scriptsize($<$0.001)} & {\scriptsize(0.026)} & {\scriptsize(1.000)} & {\scriptsize($<$0.001)} & {\scriptsize($<$0.001)} & {\scriptsize(0.091)} \\[2pt]
  Agentic-Priors & $+0.117^*$ & $+0.167^*$ & $-0.028$ & $-0.072$ & $-0.003$ & $+0.037^*$ & $-0.007^*$ \\
  & {\scriptsize($<$0.001)} & {\scriptsize(0.012)} & {\scriptsize(0.324)} & {\scriptsize(1.000)} & {\scriptsize(0.076)} & {\scriptsize(0.007)} & {\scriptsize($<$0.001)} \\[2pt]
  In-Context Learning + $z_i$ & $+0.103^*$ & $+0.264^*$ & $+0.089^*$ & $+0.065$ & $+0.470^*$ & $+0.008$ & $+0.012$ \\
  & {\scriptsize($<$0.001)} & {\scriptsize($<$0.001)} & {\scriptsize($<$0.001)} & {\scriptsize(1.000)} & {\scriptsize($<$0.001)} & {\scriptsize(1.000)} & {\scriptsize(0.091)} \\[2pt]
  HBM & $-0.100^*$ & $-0.366^*$ & $-0.193^*$ & $-0.164$ & $-0.059^*$ & $-0.006$ & $-0.123^*$ \\
  & {\scriptsize(0.036)} & {\scriptsize($<$0.001)} & {\scriptsize($<$0.001)} & {\scriptsize(0.239)} & {\scriptsize($<$0.001)} & {\scriptsize(1.000)} & {\scriptsize($<$0.001)} \\[2pt]
  HBM + $z_i$ & $-0.128^*$ & $-0.503^*$ & $-0.200^*$ & $-0.162$ & $-0.066^*$ & $-0.116^*$ & $-0.124^*$ \\
  & {\scriptsize(0.002)} & {\scriptsize($<$0.001)} & {\scriptsize($<$0.001)} & {\scriptsize(0.249)} & {\scriptsize($<$0.001)} & {\scriptsize($<$0.001)} & {\scriptsize($<$0.001)} \\[2pt]
\bottomrule
\end{tabular}
\end{table}

\FloatBarrier
\section{Prior-Scale Ablations}
\label{app::ablations}

We set the prior's covariance to $\sigma^2 I_K$ with $\sigma = 1$ rather than deriving
it from the LLM. These ablations, on Swissmetro (our development dataset), test the two
LLM-derived alternatives under the evaluation of Section~\ref{sec::exp_setup} (same test users,
posterior-predictive evaluation).

\emph{Stated uncertainty} (Table~\ref{tab::app_stated}). The LLM states a plausible range
for each quantity, which we map to a per-coefficient standard deviation. Used as given
($\times 1$), the prior is far too narrow to learn from: its log-loss barely improves with
observed choices. It must be widened about $16\times$ before it learns as fast as the
isotropic prior, at which point its cold start log-loss is much worse. No widening is
significantly better than the isotropic prior in log-loss at any $T$; in accuracy, moderate widening ($\times 4$) is
ahead after one observed choice, but not later.

\emph{Repeated elicitation} (Table~\ref{tab::app_ensemble}). Averaging eight elicitations per
user, with the same $\sigma$, is slightly worse in log-loss at every $T$. Using the
eight answers' own spread as the covariance gives a sharper cold start (higher accuracy at
$T \le 1$) but a prior too narrow to learn from: after six choices its log-loss is $0.765$
against $0.624$. The variation across repeated elicitations is thus far smaller than the
uncertainty about the user: it reflects decoding variation, not that uncertainty.

\begin{table}[h]
\centering\small
\caption{Using the LLM's stated uncertainty as the prior's spread. The LLM states a plausible range for each quantity; its implied standard deviation is multiplied by the factor shown. Swissmetro test users ($n = 100$); accuracy (\%) / log-loss after $T$ observed choices. $^*$: paired difference from the single-elicitation prior with $\sigma = 1$ (last row) significant at $p < 0.05$.}
\label{tab::app_stated}
\begin{tabular}{lcccc}
\toprule
  Prior & $T = 0$ & $T = 1$ & $T = 3$ & $T = 6$ \\
\midrule
  Stated ranges $\times1$ & 66.5 / 0.883 & 67.1 / 0.877$^*$ & 68.5 / 0.862$^*$ & 70.1 / 0.842$^*$ \\
  Stated ranges $\times4$ & 65.7 / 0.908$^*$ & 68.5$^*$ / 0.848$^*$ & 71.3 / 0.769$^*$ & 72.9 / 0.703$^*$ \\
  Stated ranges $\times8$ & 63.3 / 0.939$^*$ & 63.7 / 0.830$^*$ & 70.1 / 0.713$^*$ & 72.9 / 0.634 \\
  Stated ranges $\times16$ & 54.7$^*$ / 0.975$^*$ & 62.1$^*$ / 0.841$^*$ & 70.3 / 0.699 & 72.8 / 0.618 \\
  Stated ranges $\times32$ & 46.2$^*$ / 1.007$^*$ & 61.1$^*$ / 0.879$^*$ & 69.6 / 0.733$^*$ & 72.1 / 0.652$^*$ \\
  Stated ranges $\times64$ & 42.3$^*$ / 1.029$^*$ & 60.7$^*$ / 0.924$^*$ & 67.8 / 0.803$^*$ & 71.6 / 0.724$^*$ \\
  Single elicitation, $\sigma = 1$ (ours) & 62.5 / 0.853 & 64.7 / 0.783 & 68.8 / 0.689 & 72.4 / 0.624 \\
\bottomrule
\end{tabular}
\end{table}

\begin{table}[h]
\centering\small
\caption{Repeated elicitation: eight elicitations per user (varied phrasing, temperature and seed), averaged, with either the same $\sigma = 1$ or the eight answers' own spread as the covariance. Swissmetro test users ($n = 100$); accuracy (\%) / log-loss after $T$ observed choices. $^*$: paired difference from the single-elicitation prior with $\sigma = 1$ (last row) significant at $p < 0.05$.}
\label{tab::app_ensemble}
\begin{tabular}{lcccc}
\toprule
  Prior & $T = 0$ & $T = 1$ & $T = 3$ & $T = 6$ \\
\midrule
  8 elicitations, $\sigma = 1$ & 61.7 / 0.904$^*$ & 62.3$^*$ / 0.817$^*$ & 68.2 / 0.709$^*$ & 72.2 / 0.636$^*$ \\
  8 elicitations, own spread & 70.1$^*$ / 0.821 & 70.4$^*$ / 0.809 & 71.5 / 0.786$^*$ & 72.1 / 0.765$^*$ \\
  Single elicitation, $\sigma = 1$ (ours) & 62.5 / 0.853 & 64.7 / 0.783 & 68.8 / 0.689 & 72.4 / 0.624 \\
\bottomrule
\end{tabular}
\end{table}

\FloatBarrier
\section{Other Elicitors}
\label{app::crossmodel}

Our primary elicitor is Qwen3-32B, served privately with 4-bit quantisation and
reasoning disabled. To test sensitivity to this choice, we elicited the cold start prior with
seven open-weight models from four families through a public API, on three datasets covering
both reference scales (Swissmetro and Cars use money; Bank uses ratings). Every model
received exactly the prompts of Appendix~\ref{app::prompts}, with no per-model tuning; models
that reason by default were run with reasoning disabled. Table~\ref{tab::app_crossmodel}
reports each model as a paired difference from the primary elicitor at $T = 0$.

Accuracy is largely insensitive to the elicitor: apart from Gemma-3-12B, no model differs
significantly from the primary on Swissmetro, and on Cars and Bank the differences are within
four points in either direction. Log-loss is more sensitive, because it also penalises
miscalibrated magnitudes. The previous-generation Gemma-3-12B fails on Swissmetro and Cars,
with log-loss above that of the uninformed prior, because it states implausibly large monetary
values; its successor Gemma-4 does not.

\begin{table}[h]
\centering\small
\caption{Cold start ($T = 0$) results with other elicitors, using identical prompts and no per-model tuning (test users, $n = 100$ per dataset). First two rows: accuracy (\%) and log-loss of the primary elicitor and of the uninformed prior. Other rows: paired difference from the primary elicitor (accuracy in percentage points; positive log-loss is worse). $^*$: $p < 0.05$, paired $t$-test, unadjusted. --: not run.}
\label{tab::app_crossmodel}
\begin{tabular}{lrrrrrr}
\toprule
  Elicitor & \multicolumn{2}{c}{Swissmetro} & \multicolumn{2}{c}{Cars} & \multicolumn{2}{c}{Bank} \\
  & Acc. & LL & Acc. & LL & Acc. & LL \\
\midrule
  Primary (Qwen3-32B, 4-bit, private) & 62.5 & 0.853 & 52.3 & 0.989 & 62.7 & 0.672 \\
  Uninformed prior & 25.8 & 1.092 & 33.6 & 1.101 & 51.0 & 0.693 \\
\midrule
  \texttt{mistralai/mistral-small-24b-instruct-2501} & $-0.2$ & $+0.017$ & $-2.9$$^*$ & $+0.033$$^*$ & $+1.7$ & $-0.005$$^*$ \\
  \texttt{qwen/qwen3-30b-a3b-instruct-2507} & $+0.5$ & $+0.064$$^*$ & $-0.3$ & $-0.005$ & $-1.5$ & $+0.002$ \\
  \texttt{qwen/qwen3.8-27b} & $-0.9$ & $+0.043$$^*$ & -- & -- & -- & -- \\
  \texttt{google/gemma-4-31b-it} & $+0.2$ & $+0.029$$^*$ & $-1.1$ & $+0.004$ & $+2.5$$^*$ & $-0.006$$^*$ \\
  \texttt{google/gemma-4-26b-a4b-it} & $-0.1$ & $+0.039$$^*$ & $-2.2$ & $+0.013$$^*$ & $+2.1$$^*$ & $-0.006$$^*$ \\
  \texttt{meta-llama/llama-3.1-8b-instruct} & $-1.9$ & $+0.100$$^*$ & $-0.8$ & $+0.015$$^*$ & $-3.9$$^*$ & $-0.000$ \\
  \texttt{google/gemma-3-12b-it} & $-5.1$ & $+1.097$$^*$ & $-6.1$$^*$ & $+0.152$$^*$ & $+0.9$ & $-0.003$ \\
\bottomrule
\end{tabular}
\end{table}

\FloatBarrier
\section{Diagnostics}
\label{app::diagnostics}

These analyses explain the results of Section~\ref{sec::results} from Agentic-Priors' side. Both use the same runs:
Agentic-Priors + $z_i$ and the HBM without $z_i$, on the same test users, task permutations
and posterior-draw seeds, so every difference is paired.

\emph{Covariance swap} (Table~\ref{tab::app_covswap}). Each prior is split into its mean and its
covariance, and the two halves are exchanged between Agentic-Priors and the HBM. At $T = 0$,
exchanging covariances barely moves either prior: with the HBM's covariance, our prior's accuracy
changes by at most 2 points on every dataset, and the HBM with our covariance stays
within about 1 point of the HBM. The cold start gap therefore lies in the prior's mean. Once
choices are observed, learning speed follows the covariance: with the HBM's covariance, our mean
gains more from the observed choices on every dataset but Swissmetro, and after the last step all
users reach it is within about 1 point of the HBM's accuracy on LPMC, Cars CN, Beer and Bank.
Conversely, the HBM's mean with our isotropic covariance learns more slowly than the HBM on most
datasets.

\emph{What a small survey adds} (Table~\ref{tab::app_hybrid}). Our mean is combined with the
covariance of an HBM fitted on $N$ surveyed users and compared with that HBM, using the
exchange-rate surveys of Section~\ref{sec::res_exchange}. In log-loss, our mean is better only for the smallest
surveys (EV at $N = 10$, Beer up to $N = 15$); from 25 users the survey's own mean is as
good or better on every dataset, and in accuracy it is better at every $N$ on six datasets. A
small survey thus estimates the mean the LLM must guess, which is why the prior's exchange
rate is small.

\begin{table}[h]
\centering\small
\caption{Covariance swap between Agentic-Priors + $z_i$ (ours) and the HBM without $z_i$: accuracy (\%) / log-loss of each prior, of our mean with the HBM's covariance, and of the HBM's mean with our covariance $\sigma^2 I$ (test users, $n = 100$ per dataset).}
\label{tab::app_covswap}
\setlength{\tabcolsep}{4pt}
\begin{tabular}{lrrrr}
\toprule
  & Ours & Ours + HBM cov. & HBM + our cov. & HBM \\
\midrule
\multicolumn{5}{l}{\emph{$T = 0$}} \\
  Swissmetro & 62.5 / 0.853 & 64.3 / 0.898 & 68.8 / 0.845 & 68.5 / 0.753 \\
  LPMC & 61.9 / 1.128 & 60.8 / 1.208 & 72.3 / 0.915 & 71.1 / 0.762 \\
  Cars & 52.3 / 0.989 & 53.4 / 1.010 & 64.6 / 0.806 & 64.0 / 0.797 \\
  Cars CN & 43.6 / 1.150 & 43.4 / 1.159 & 49.9 / 0.998 & 50.5 / 0.985 \\
  EV & 53.5 / 0.689 & 54.7 / 0.687 & 64.9 / 0.631 & 65.8 / 0.629 \\
  Beer & 48.2 / 1.048 & 47.1 / 1.061 & 45.0 / 1.107 & 46.1 / 1.042 \\
  Bank & 62.7 / 0.672 & 63.7 / 0.669 & 74.3 / 0.562 & 74.1 / 0.549 \\
\midrule
\multicolumn{5}{l}{\emph{Last step all users reach}} \\
  Swissmetro ($T = 6$) & 72.4 / 0.624 & 73.8 / 0.615 & 76.5 / 0.547 & 77.0 / 0.561 \\
  LPMC ($T = 3$) & 81.1 / 0.637 & 83.3 / 0.721 & 82.9 / 0.550 & 83.9 / 0.504 \\
  Cars ($T = 12$) & 67.8 / 0.764 & 70.3 / 0.724 & 71.7 / 0.701 & 72.8 / 0.659 \\
  Cars CN ($T = 12$) & 54.2 / 1.006 & 54.7 / 1.024 & 54.5 / 0.964 & 55.3 / 0.923 \\
  EV ($T = 3$) & 73.3 / 0.551 & 75.2 / 0.539 & 77.8 / 0.478 & 78.6 / 0.460 \\
  Beer ($T = 5$) & 73.6 / 0.666 & 78.4 / 0.615 & 75.0 / 0.646 & 78.3 / 0.584 \\
  Bank ($T = 11$) & 70.9 / 0.590 & 77.5 / 0.523 & 75.4 / 0.547 & 77.9 / 0.483 \\
\bottomrule
\end{tabular}
\end{table}

\begin{table}[h]
\centering\small
\caption{What a survey of $N$ users adds to the LLM's mean: our prior's mean combined with the covariance of an HBM without $z_i$ fitted on $N$ users, minus that HBM, at $T = 0$ (positive: our mean is better; averaged over survey draws; $^*$: $p < 0.05$, paired, unadjusted; --: $N$ not below the pool).}
\label{tab::app_hybrid}
\setlength{\tabcolsep}{3pt}
\begin{tabular}{lrrrrrrrr}
\toprule
  & $N = 10$ & $N = 15$ & $N = 25$ & $N = 50$ & $N = 100$ & $N = 200$ & $N = 400$ & All \\
\midrule
\multicolumn{9}{l}{\emph{Log-loss}} \\
  Swissmetro & $-0.088^*$ & $-0.117^*$ & $-0.127^*$ & $-0.146^*$ & $-0.137^*$ & $-0.137^*$ & $-0.139^*$ & $-0.145^*$ \\
  LPMC & $+0.208$ & $-0.018$ & $-0.297^*$ & $-0.380^*$ & $-0.353^*$ & $-0.380^*$ & $-0.430^*$ & $-0.446^*$ \\
  Cars & $-0.170^*$ & $-0.197^*$ & $-0.217^*$ & $-0.217^*$ & $-0.214^*$ & $-0.217^*$ & -- & $-0.214^*$ \\
  Cars CN & $-0.089$ & $-0.110$ & $-0.129$ & $-0.146^*$ & $-0.154^*$ & $-0.164^*$ & -- & $-0.174^*$ \\
  EV & $+0.047^*$ & $+0.004$ & $-0.014$ & $-0.044^*$ & $-0.046^*$ & $-0.055^*$ & $-0.055^*$ & $-0.058^*$ \\
  Beer & $+0.110^*$ & $+0.063^*$ & $+0.016$ & $-0.009$ & $-0.014$ & $-0.014$ & $-0.019$ & $-0.019$ \\
  Bank & $+0.061$ & $-0.078^*$ & $-0.092^*$ & $-0.116^*$ & $-0.124^*$ & $-0.121^*$ & $-0.120^*$ & $-0.120^*$ \\
\midrule
\multicolumn{9}{l}{\emph{Accuracy (percentage points)}} \\
  Swissmetro & $-6.7^*$ & $-8.6^*$ & $-8.3^*$ & $-9.9^*$ & $-5.5^*$ & $-7.0^*$ & $-5.2^*$ & $-4.1$ \\
  LPMC & $-6.9^*$ & $-6.5^*$ & $-5.9^*$ & $-8.3^*$ & $-8.5^*$ & $-9.6^*$ & $-11.6^*$ & $-10.3^*$ \\
  Cars & $-14.8^*$ & $-14.9^*$ & $-14.8^*$ & $-13.4^*$ & $-12.9^*$ & $-12.6^*$ & -- & $-10.6^*$ \\
  Cars CN & $-6.0^*$ & $-6.6^*$ & $-7.1^*$ & $-7.6^*$ & $-6.8^*$ & $-7.3^*$ & -- & $-7.1^*$ \\
  EV & $-6.5^*$ & $-7.9^*$ & $-7.7^*$ & $-9.2^*$ & $-8.6^*$ & $-11.3^*$ & $-10.8^*$ & $-11.1^*$ \\
  Beer & $+3.6^*$ & $+2.4$ & $+1.8$ & $+2.0$ & $+1.2$ & $+2.2$ & $+0.7$ & $+1.0$ \\
  Bank & $-12.1^*$ & $-13.8^*$ & $-13.8^*$ & $-14.2^*$ & $-13.2^*$ & $-12.3^*$ & $-12.9^*$ & $-10.4^*$ \\
\bottomrule
\end{tabular}
\end{table}

\FloatBarrier
\section{Exchange-Rate Method}
\label{app::exchange}

The exchange rate expresses the value of Agentic-Priors + $z_i$ as the number of surveyed
users whose data it is worth. For each dataset, we fit HBM + $z_i$ (as in the
benchmark of Section~\ref{sec::benchmarks}, with the same prompt-matched user characteristics and settings) to random
\emph{surveys} of $N$ users drawn from the non-test users, and score each fit at
$T = 0$ on the same test users, task permutations and posterior-draw seeds as our prior.

\emph{Surveys.} $N$ runs over 2--9, 10, 15, 25, 50, 100, 200 and 400 (below the pool size), plus
the whole pool. Each $N$ is drawn 10 times for $N \le 50$ and 3 times above; each draw is
nested, so a draw's survey of size $N$ contains its smaller surveys. \texttt{bayesm} rejects some
of the smallest surveys: LPMC's when a travel mode is never chosen, EV's when the covariance
cannot be estimated. Rejected fits are skipped, so LPMC and EV cannot be fitted below 6 and 3
users (Table~\ref{tab::app_exchange}).

\emph{Exchange rate.} For each $N$, the paired difference between the HBM and our prior is
averaged over users and survey draws. $N^*$ is where a non-decreasing least-squares fit of
this difference, against $\log N$, crosses zero (interpolated in $\log N$); its 95\% interval
comes from 2{,}000 bootstrap resamples of the test users. When the HBM is already ahead at
the smallest fittable survey, only a bound is reported. In Figure~\ref{fig::exchange}, each $N$'s band
combines the user and survey-draw uncertainty of that $N$; HBMs fitted on the smallest
surveys can be badly over-confident, which is why our prior can be worth a few users even
where it is worse than the uninformed prior (Cars CN, log-loss).

\begin{table}[h]
\centering\small
\caption{Exchange rate $N^*$ of Agentic-Priors + $z_i$ against HBM + $z_i$: the survey size at which the HBM overtakes the prior, with its 95\% user-bootstrap interval. $<$$n$: the HBM is already ahead at the smallest survey that could be fitted. Smallest fit: smallest $N$ with at least one successful fit; rejected: fits bayesm rejected across all $N$ and draws; pool: non-test users.}
\label{tab::app_exchange}
\begin{tabular}{lllrrr}
\toprule
  & Log-loss $N^*$ & Accuracy $N^*$ & Smallest fit & Rejected & Pool \\
\midrule
  Swissmetro & 9.6 [5.8, 23] & 5.9 [5.0, 19] & 2 & 12 & 1091 \\
  LPMC & 18 [13, 21] & $<$6 [$<$6, 6.2] & 6 & 82 & 1204 \\
  Cars & 7.1 [5.5, 8.4] & $<$2 [$<$2, 3.2] & 2 & 1 & 284 \\
  Cars CN & 5.7 [3.9, 8.2] & 3.3 [$<$2, 10.0] & 2 & 0 & 348 \\
  EV & 18 [13, 33] & $<$3 [$<$3, 6.1] & 3 & 24 & 883 \\
  Beer & 38 [23, 65] & 65 [19, 432] & 2 & 8 & 1191 \\
  Bank & 12 [9.4, 14] & 4.0 [$<$2, 6.7] & 2 & 9 & 846 \\
\bottomrule
\end{tabular}
\end{table}


\end{document}